# First-principles modeling of the interactions of iron impurities with graphene and graphite

Danil W. Boukhvalov[*,1]

[1]Computational Materials Science Center, National Institute for Materials Science, 1-2-1 Sengen, Tsukuba, Ibaraki 305-0047, Japan

*Results of first principles modelling of interactions graphene and graphite with iron impurities predict the colossal difference between these two carbon allotropes. Insertion of the iron atoms between the planes of graphite is much more energetically favourable than adsorption of the iron adatom at graphite or graphene surface. High mobility of iron adatom over graphite surface and within bulk graphite is reported. Calculated values of formation energies for the substitutional iron impurities suggest that iron is more destructive for graphite than for graphene. This effect caused formation of uniform carbon environment of the iron atom inside the multilayer system. In contrast to graphene segregation of the substitutional iron impurities in graphite at the ambient conditions is not energetically favourable. Enhancement of interlayer bonding in contaminated graphite and purity of graphene from iron impurities are also reported.*

---

\* Corresponding author: e-mail BUKHVALOV.Danil@nims.go.jp,

**1 Introduction** Since discovery of the magnetism in pure [1] and proton irradiated [2, 3] graphite and nanographites [4] possible role of transition metals impurities in magnetic carbon [5] compounds are evergreen topic. The most propagated magnetic impurity in carbon systems is the iron [6]. Recent XMCD measurements [7, 8] and direct measurements of amount of the impurities in graphite [2] suggests for neglected number of the iron and other transition metal impurities and intrinsic nature of the magnetism in graphite. Experimental results for Fe-bombed graphite [9] evidences for the uniform distribution of iron paramagnetic impurities without formation of large clusters or magnetic two and three dimensional structures. Why iron adatoms attendant in graphene as natural impurities or extrinsic defects prefer dissipation instead segregation is the challenge for theoretical investigations.

Production of the one of the most prospective novel electronic material known as graphene from graphite [10] require studying of possible presence of transition metals impurities at graphene sheets or as experimentally observed substitutional defects [11]. Discovery of several graphene physical properties in graphite [12, 13] needs to explore the possible disposition of the impurities in graphite because graphene is extremely sensitive to the each single adsorbed specie [14]. Cutting of graphene by transition metals adatoms and clusters had been observed experimentally [15]-[17] and discussed theoretically [18] is the challenge for explore the role of the possible destruction of graphite for the formation of nano-graphites with given shape [19].

In present work I report the results of the modeling for placement of the single iron adatom at the carbon scaffold and the energetics of migration of iron impurities. Also the modelling of possible clusterization of substitutional iron atoms in graphite, surface and bulk layers of graphite and its magnetic properties are described.

**2 Computational method and model** Density functional theory (DFT) calculations have been carried out with the same pseudopotential code SIESTA [19] used for our previous modeling of functionalized graphene multilayers and graphite [18,21]-[23]. All calculations were carried out for energy mesh cut off 360 Ry and k-point mesh 4×4×2 in Mokhorst-Park scheme [24]. During the optimization, the electronic ground state was found self-consistently using norm-conserving pseudo-potentials for cores and a double-ζ plus polarization basis of localized orbitals for carbon and iron. Optimization of the atomic forces and total energies was performed with an accuracy of 0.04 eV/Å and 1 meV, respectively.

The minimal model of graphite surface is graphene six layers slab. The number of layers was chosen with taking into account the measurements of Raman spectras of graphene multylayers and graphite [25]. For the modeling of bulk graphite interaction with iron atom is inserted between the central layers of the same six layers graphene slab (see Fig. 1a). For the proper separation of studied impurities within periodic conditions the supercell 3×3×1 has been used. The distance between equivalent adatoms for this supercell is about 1 nm (four lattice parameters). The total number of carbon atoms in used supercell is 192.

Formation energy have been calculated within standard formula $E_{form} = E_{graphite + iron} - (E_{carbon} \cdot N + E_{iron})$ where $E_{carbon}$ is the energy per carbon atom in pristine graphite supercell, N is the total number of carbon atoms in studied supercell and $E_{iron}$ is the energy of single iron atom in the empty box. The calculations of the binding energies between layers have been performed with the formula $E_{bind} = 5/6 E_{graphite} + E_{graphene\ monolayer + iron} - E_{graphite + iron}$. Calculated binding energy between layers in 6 layer graphite slab is 26 meV per carbon atom that is rather close to same value for the bulk graphite (33 meV/C) [20]. Interlayer distances in six layer slab are also almost the same as in bulk graphite.

Main issue of the calculation of adsorption on graphite is the overestimation of covalent [22] and strong ionic bonds within LDA [26] approach and underestimation of weak bonds (such as van der Waals or π-π) within GGA [27]. Calculation of multilayers graphitic systems with using GGA provide significant spread between layers (see discussions in Ref. [21] and reference therein). The solution of this problem require the calculation of single iron adsorption on graphene monolayer within LDA and compare the results with previous GGA calculations [18, 28, 29]. The formation energy differences for various iron positions on graphene monolayer (Fig. 1) calculated with using LDA and GGA functionals ($E^0_{formLDA} - E^0_{formGGA}$) is close to the same value. For obtain the correct values of formation energies for adsorbed and substitutional atom of iron on graphene and graphite it is necessary to shift calculated within LDA value of the energy to described above energy difference ($E_{form} = E_{formLDA} - (E^0_{formLDA} - E^0_{fofmGGA})$). All discussed below energies have been calculated by described method.

**3 Adsorption of iron adatom on graphite** The calculations for the three more favorable positions [28] had been performed for obtain the most energetically favourable position of iron impurity over graphene, graphite and between graphite layers (Fig. 2a). For the case of graphite surface the most stable position of the single iron adatom is over the carbon atom (Fig. 1a) in contrast to graphene monolayer where iron adatoms prefer to sit over hollow center of the carbon hexagon. This difference caused by the less flexibility of the top layer of graphite which is also dramatically changes the energetics of chemisorption process [22]. But the values of binding energies for graphite surface and graphene are very close that cased weak interactions between layers.

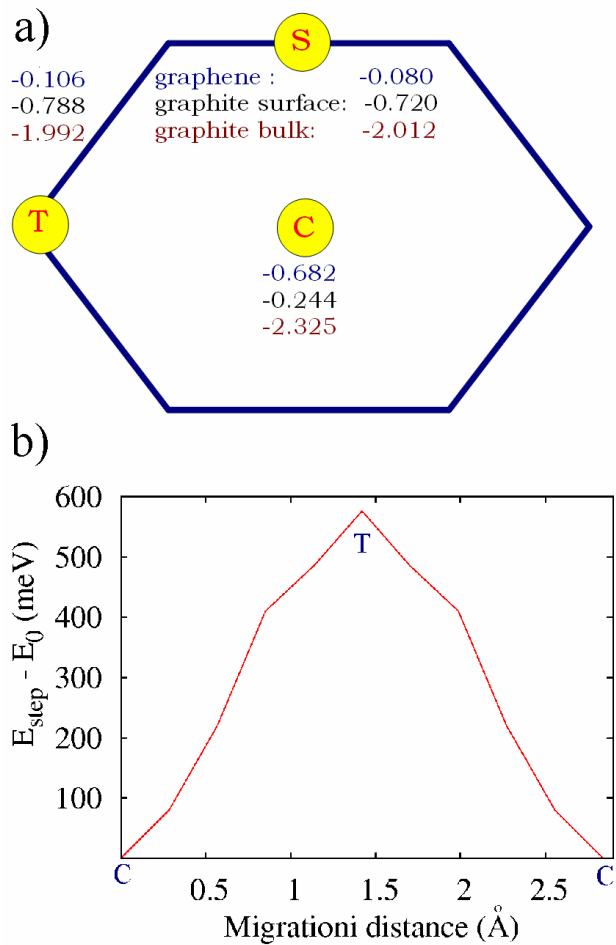

**Figure 1** (a) A sketch of the iron adatom (yellow circle) disposition over the centre carbon hexagon (C), carbon atom (T) and bridge between two carbon atoms (S) with appropriate formation energies (in eV). (b) Migration barrier for the migration of the iron adatom over graphene monolayer from C via T to other C position.

Energy of the migration has been defined as the difference between the formation energies for the most stable configuration (C for graphene and T for graphite surface) and the next less stable configuration (T for graphene and S for graphite surface). For check the migration barrier for the move of the iron adatom over graphene monolayer the energetics of the migration from C to T and to another C position have been examined. How we can conclude from the results of calculations (see Fig. 1b) preformed with using the method described in Ref. [29] discussed above energy difference between the most stable and the next less stable configurations is the energy barrier. For graphene the value of the energy barrier of migration is 0.576 eV which is about six time higher that the binding energy between graphene and iron adatom at the intermediate step of migration. In contrast to graphene the energy required for iron adatom migrated over graphite surface is 0.068 eV and about ten times smaller that the binding energy of impurity with carbon host. The magnetic moment of the iron adatom over the hollow of the graphene hexagon is about 2 $\mu_B$ which is in agreement with previous studies [29, 30]. The presence of the additional layers of graphite dramatically changes the distribution of the *3d* electrons of the iron caused by the different carbon environments. The magnetic moment there is about 3 $\mu_B$ for the centre of the hexagon (C), and 4.8 $\mu_B$ for the T and S positions (see Fig. 1). When iron adatom placed over the hollow site (C) there is $z^2$ and $x^2-y^2$ orbitals occupied for both spin, but for the S and T position only $z^2$.

In contrast to the graphite surface being of the iron atom between two graphenic sheets lead of connection with carbon atoms from the top and bottom (Fig. 2a). Uniform carbon environment is more favorable for transition metal adatom and results the significant diminishment of formation energy (Fig. 1). The magnetic moment of single iron impurity between graphite layers is about 3.0, 3.2 and 3.3 $\mu_B$ for C, S and T position respectively. The migration energy for the iron atom in bulk graphite is 0.313 eV that is much smaller than binding energy. Calculated small value of migration energy is evidence of high fluidity of iron adatoms between graphene planes.

Lower formation energy of iron adatom provides enhancement of interlayer binding energy from 26 meV/C for the pure graphite slab to 94 meV/C for graphite slab with encapsulating iron adatom. Binding energy of the top layer with bulk graphite increase insignificantly from 26 to 28 meV/C for the case of iron adatom adsorption over graphite.

**4 Substitutional iron adatoms** For the case of substitutional impurity the difference between graphene and graphite are also significant. For graphene monolayer small shift of substitutional impurity out of graphene plane have been found in recent theoretical works [18, 31]. The presence of the second carbon layer creates the opportunity for substitutional iron atom to obtain the uniform environment of light elements. It leads more significant shift of the iron atom into the interlayer area (Fig. 2d). Similar to the case of adatom between graphenic planes of graphite iron adatom formation of uniform carbon environment provides significant diminishment of the formation energy. Similarly to iron adatom in bulk graphite significant increase from 26 to 116 meV/C of interlayer binding energy between layer with substitutional iron adatom and graphite have been calculated. The values of formation energies for the single substitutional iron impurity are 3.72, 0.81 and 0.84 eV for the cases of graphene, graphite surface and bulk respectively.

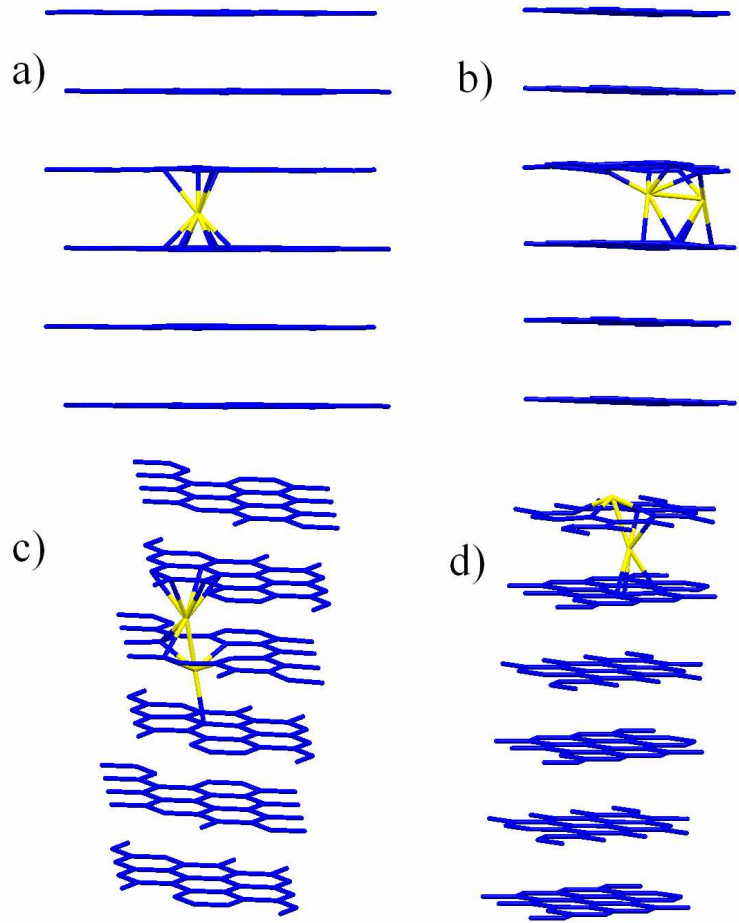

**Figure 2** Optimized atomic structure for iron adatom (shown by yellow) inserted between graphite (shown by blue) layers (a), two distant substitutional iron impurities in bulk graphite (b), two substitutional iron impurities in the nearest positions in bulk graphite (c) and graphite surface (d).

Studying of the clusterization processes of substitutional iron impurities require the calculation of the formation energies for various distances. For graphene monolayer light tendency to clusterization have been observed (Fig. 3) due to decay of formation energy with the approach iron atoms to each other. For the bulk and surface of graphite decreasing of distance between remote substitutional impurities does not provide the diminishment of the formation energy except the case of pair formation for the nearest substitutional impurities. It needs to note that several intermediate positions of migration (for example near 5 on Fig. 3) the energy cost about 0.4 eV are added to the ordinary barrier of migration. In contrast to the distant substitutional iron adatom (Fig. 2b) formation of pairs leads to the shift one of the iron atoms up to carbon flat and second down (Fig. 2c, d). Segregation of substitutional iron impurities in graphite with the possible formation of iron pairs is very energetically favourable but intermediate steps of migration of iron atoms to each other is endothermic and clusterization of iron in graphite require annealing of the samples. Magnetic moment of single substitutional iron atom in graphene monolayer and in the surface layer of graphite is about 2.0 $\mu_B$, and 2.8 $\mu_B$ in the central layer of graphite. In contrast to the previous differences between three studied systems exchange interactions in its entire are rather similar. The energy differences between ferromagnetic and antiferromagnetic configurations for the two substitutional iron impurities for distances above 4 Å suggest for RKKY-like magnetic interactions similarly to the substitutional cobalt impurity in graphene [32]. The value of exchange interactions are decay from 30 meV for the distance about 5 Å to 10 meV for 10 Å. For the distances smaller than 4 Å magnetic interactions switch from antiferromagnetic to ferromagnetic with grow of energy difference to 100 meV similarly to the ferromagnetism of the iron carbide (cementite). Formation of the iron dimmer (Fig. 2c, d) provides further changes of magnetic properties. The Magnetic moments is about 2.9 $\mu_B$ for all systems and exchange turn to antiferromagnetic with the difference between both magnetic configuration 297 meV for graphene, 430 for graphite surface and 300 meV for the central layer of graphite. Discussed changes of magnetic properties suggest for the possibility of ferromagnetic cementite-like phase formation only in graphene. In graphene this phase will be unstable due to the segregation of the iron clusters. Obtained results also explain experimentally detected presence only paramagnetic ions in iron-irradiated graphite without formation of magnetic web or clusterization of impurities at ambient conditions [9].

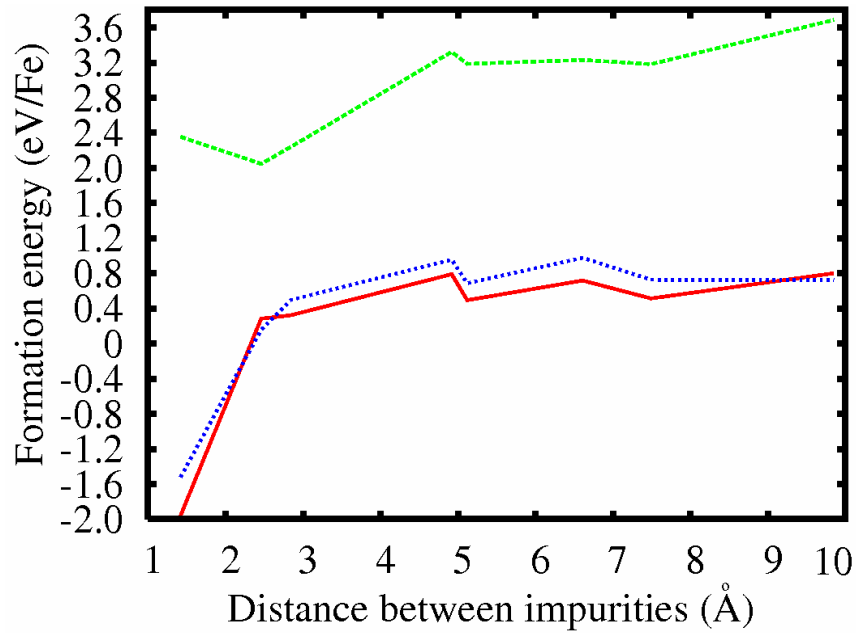

**Figure 3** Calculated formation energy per iron impurity for the substitution of the distant carbon atom in bulk graphite (solid red line), graphite surface (dotted blue line), and graphene monolayer (dashed green line).

For the case of nearly placed iron adatoms diminishment of the formation energy of substitutional iron atoms and pairs of atoms from graphene monolayer to graphite evidences that iron is more destructive for graphite than for graphene due to substitution of carbon atoms. This situation could be realized for the high concentration of iron impurities or for the presence the iron particles inside graphite. Described effect could be used for the production of new magnetic materials. Zig-zag graphene edges are in perspective for the further electronic applications [33, 34]. Unfortunately the reconstructions of magnetic edges caused by the self-passivation of dangling bonds [35]-[38] or the oxidation of its [23, 39]-[41] switch off the magnetism on the edges of graphene monolayers. In absence of oxidative species only graphite magnetic zig-zag edges is stable [42]. The methods of non-oxidative cut of graphene by metal nanoparticles [15]-[17] could be suitable for the formation of nano-graphites with given shape and stable magnetic edges because the lower formation energies of substitutional iron adatoms in graphite.

**5 Conclusions** Performed DFT modeling of the interaction of iron adatoms with graphite argue for the high mobility of interstitial atoms in bulk graphite and adatoms over graphite surface. The energy barrier value for this type of migration is smaller than 0.3 eV that make possible discussed migration at ambient conditions [11, 44, 45].

For the case of substitution of carbon atom by iron the colossal differences between graphene and graphite have been found. For the case of graphene the pairing of substitutional adatoms is the very energetically favorable. For the bulk and surface of graphite forthcoming of the remote iron atoms is not energy favourable. Another crucial difference between graphene and graphite is rather smaller formation energies for the substitutional iron adatoms. Calculated diminishments of the formation energy are caused by the shift of substitutional impurity to the interlayer space and restore of its uniform carbon environment. Obtained results suggest for the production of nanographites with a given shape due to nonoxidative cut by transition metals nanoparticles.

Significant grow of interlayer binding energy in the presence of iron impurity makes delamination [10, 43] of contaminated layers energetically unfavorable. Instability of iron adatom on graphene surface is also suggested for the purity of graphene from contamination by iron impurities unavoidable for graphite.